\def\be{\begin{equation}}
	\def\ee{\end{equation}}
\def\ba{\begin{array}}
	\def\ea{\end{array}}

\documentclass[prl,showpacs,twocolumn,amsmath]{revtex4}
\usepackage{amsmath}
\usepackage{amsfonts}
\usepackage{mathrsfs}
\usepackage{amssymb}
\usepackage{pifont}
\usepackage{epsfig,subfigure,dsfont,amsthm,amsbsy,mathrsfs,amscd}
\usepackage{epstopdf}
\usepackage{bbm}
\usepackage{color}     
\def\qed{\leavevmode\unskip\penalty9999 \hbox{}\nobreak\hfil
	\quad\hbox{\leavevmode  \hbox to.77778em{%
			\hfil\vrule   \vbox to.675em%
			{\hrule width.6em\vfil\hrule}\vrule\hfil}}
	\par\vskip3pt}
\usepackage{leftidx}

\begin{document}
	\title{\large\bf  The Ergotropy of Quantum Batteries under Unruh Effect}
	\author{Yiding Wang, Shengyan Ma, Xiaofen Huang and Tinggui Zhang$^{\dag}$}
	\affiliation{ School of Mathematics and Statistics, Hainan Normal University, Haikou, 571158, China \\
		$^{\dag}$ Correspondence to tinggui333@163.com}
	
	\bigskip

	\bigskip
	\bigskip
	
	\begin{abstract}
	We study the effects of uniform acceleration on the ergotropy of quantum batteries modeled as Unruh-DeWitt detectors, in both bipartite and tripartite setups. In the bipartite system, we systematically compare three scenarios: accelerating the battery, accelerating the charger, and accelerating both simultaneously. We find that only battery acceleration can induce a sudden emergence of ergotropy at a critical acceleration threshold, while the corresponding composite-system energy change may increase or decrease depending on the initial state at the onset of acceleration. Charger acceleration leaves the battery ergotropy constant within the perturbative regime, while simultaneous acceleration leads to monotonic decay due to coherent cancellation of the $q$-dependence. Extending to a tripartite system with one battery and two chargers, we find that battery acceleration again induces ergotropy emergence, whereas accelerating adjacent charger does not—consistent with the bipartite charger-acceleration case. These results reveal that the Unruh effect plays a dual role, both enhancing and degrading quantum battery performance, depending on which subsystem is accelerated and on the multipartite structure, bridging relativistic quantum field theory and quantum thermodynamics with relevance to experimentally accessible platforms.
	\end{abstract}
	
	 \maketitle
	
	\section{I. Introduction}
   The potential for developing energy technologies within the quantum regime has long been a central theme in quantum thermodynamics and quantum physics \cite{tdk,kvhm,rual}. Over the past decade, quantum batteries—quantum systems that store and release energy by operating on quantum-mechanical principles—have emerged as a promising framework for future energy technologies \cite{mplk,fcfa,dfmc,gmam,jmmf}. The concept of quantum batteries was first introduced by Alicki and Fannes \cite{ramf}. This pioneering work demonstrated that quantum resources, such as entanglement and coherence, play a key role in the process of efficient work extraction.
   
   Since its proposal, the concept of quantum batteries has attracted considerable research interest. Unlike traditional batteries that rely on classical electrochemical processes, quantum batteries can achieve superior performance by harnessing quantum effects and collective phenomena \cite{mplk,amar,lpga}. Research on quantum battery models has also attracted significant attention, leading to the proposal of various effective theoretical frameworks such as multipartite spin-chain models \cite{tplj,drgm,sgtc,fqdh,rgds}, Tavis-Cummings models \cite{gmam,jxlh}, harmonic oscillator models \cite{gmad,dfgm,yvda}, Dicke models \cite{lfmp,yyzt,fqdy}, Sachdev-Ye-Kitaev models \cite{drdr,drgma,fdjm}, and cavity-QED architectures \cite{lwsq,mlht}. Furthermore, since any quantum system inevitably interacts with its environment, the performance of quantum batteries in dissipative settings has also attracted significant research interest \cite{wctr,mbah,lwsql,wlsh,sqll}. For a comprehensive review of these quantum battery models and related research, see Ref. \cite{fcsg}.
   
   Spin-chain quantum batteries, as an important class of many-body quantum battery models, have attracted considerable attention in recent years \cite{aasa,pysh,sczz}. In Ref. \cite{aasa}, the authors investigated the effects of Dzyaloshinsky–Moriya and KSEA interactions, temperature, and quantum coherence on the ergotropy and battery capacity in a Heisenberg spin-chain quantum battery. Sun et al. proposed a cavity–Heisenberg spin-chain quantum battery model and found that increasing the spin size significantly enhances charging energy and power, while cavity–spin entanglement plays opposite roles in closed versus open systems \cite{pysh}. On the other hand, significant progress has been made in the research field of solid-state quantum batteries \cite{rgds,lrgg,ckhc}. Grazi et al. in \cite{rgds} studied a one-dimensional dimerized XY spin-chain quantum battery and found that when the charging quench crosses a quantum phase transition, the stored energy exhibits robustness against variations in charging time and parameters in the thermodynamic limit, providing a theoretical basis for designing stable solid-state quantum batteries. A cyclic solid-state quantum battery model based on two interacting qubits has also been proposed recently \cite{lrgg}, which charges via a thermal bath and extracts ergotropy after decoupling the subsystems, and has been successfully simulated and verified on IBM superconducting quantum hardware, achieving an operating mechanism with efficiency exceeding 50\% and finite ergotropy. Hu et al. \cite{ckhc} experimentally realized a solid-state quantum battery with 2 to 12 Transmon qubits on a superconducting quantum processor, demonstrating that significant quantum charging advantages can be achieved with only nearest-neighbor two-body interactions, and measured quantum features such as coherent and incoherent ergotropy, as well as entanglement. It is worth noting that the experimental platforms mentioned above, including superconducting qubits and spin chains, have two-level systems as their core building blocks, which is precisely the basic assumption of the Unruh‑DeWitt detector model. By introducing the Unruh effect into this universal quantum battery framework, our work bridges relativistic quantum field theory and quantum thermodynamics, while revealing the influence of acceleration as an unconventional control knob on quantum energy storage. Thus, the contribution of this paper is not merely a purely relativistic theoretical exploration, but also relevant to the broader physics of quantum batteries.
   
   In the past two years, increasing attention has been paid to the convergence of quantum thermodynamics and curved spacetime quantum field theory, particularly regarding the performance of quantum batteries in curved spacetime \cite{xhky,jlxc,amsg,ztxl,xlzt,xlzt2,ycww,zlyl}. Compared with recent studies on relativistic quantum batteries, the novelty of our work lies in the following aspects. First, most existing works treat the battery as a whole, without distinguishing the battery subsystem from the charger subsystem. Here, we explicitly separate their roles and systematically compare the asymmetric effects of accelerating the battery versus accelerating the charger. Second, prior works have largely focused on two‑body systems, with little attention to many‑body structures in relativistic quantum batteries. We extend the analysis to three‑body systems and uncover new phenomena arising from the extra degrees of freedom, such as charger‑induced emergence of ergotropy. These results provide new insights into the influence of acceleration on quantum batteries and offer a preliminary exploration of the novel challenges and potential control possibilities that relativistic effects may pose for the practical performance of quantum thermodynamic devices.
   
   The rest of this paper is organized as follows. In Section II, we study, within a bipartite system, the effects of a uniformly accelerated battery, a uniformly accelerated charger, and both accelerating simultaneously (corresponding to Scenario 1, Scenario 2, and Scenario 3, respectively) on the ergotropy of battery subsystem. In Section III, we extend our investigation to a tripartite quantum system composed of a battery and two chargers. Here, we examine the influence of a uniformly accelerated battery, a uniformly accelerated adjacent charger, and the battery and the nearby charger accelerate simultaneously (corresponding to Scenario 1, Scenario 2, and Scenario 3) on the ergotropy of the battery subsystem. We summarize and discuss our conclusions in the last section.

	\section{II. Unruh Effect on bipartite Quantum Batteries}
	In this work, we select ergotropy \cite{ramf,aear,assd,rpas} as a key figure of merit for evaluating the performance of quantum batteries. It is defined as the maximum energy that can be extracted via unitary operations, expressed as:
	\begin{equation}\label{e6}
		\varepsilon(\rho)=E(\rho)-\min\limits_{U}\text{Tr}(HU\rho U^\dagger),
	\end{equation}
	where $E(\rho)=\text{Tr}(H\rho)$ is the energy of battery state $\rho$, $H$ denotes the battery system Hamiltonian, and $U$ represents the unitary operation acting on the battery state. By rearranging the eigenvalues of the density matrix $\rho$ in descending order and those of the Hamiltonian $H$ in ascending order, i.e.,
	\begin{equation*}
		\begin{split}
			&\rho=\sum_{i=0}^{d-1}\lambda_i|\lambda_i\rangle\langle\lambda_i|,\,\lambda_0\geq\lambda_1\geq\cdots\geq\lambda_{d-1},\\
			&H=\sum_{i=0}^{d-1}\epsilon_i|\epsilon_i\rangle\langle\epsilon_i|,\,\epsilon_0\leq\epsilon_1\leq\cdots\leq\epsilon_{d-1},
		\end{split}
	\end{equation*}
	the corresponding passive state can be written as
	\begin{equation*}
	\pi_\rho=\sum_{i=0}^{d-1}\lambda_i\epsilon_i.
	\end{equation*}
	Based on this concept, the definition of ergotropy given by Eq. (\ref{e6}) can be further simplified to
	\begin{equation}\label{e7}
		\varepsilon(\rho)=E(\rho)-E(\pi_\rho)=E(\rho)-\sum_{i=0}^{d-1}\lambda_i\epsilon_i.
	\end{equation}

	In this section, we consider a quantum battery model consisting of a two‑level battery coupled to a two‑level charger via anisotropic Heisenberg coupling. To clearly describe the occupation and transfer of excitations, we employ fermionic creation and annihilation operators to model the system. The total Hamiltonian of the system can be written as:
	\begin{equation*}
	\begin{split}
	H&=H_b+H_c+f(t)H_I=\kappa_bc_b^\dagger c_b+\kappa_cc_c^\dagger c_c+f(t)H_I,
	\end{split}
	\end{equation*}
	with the interaction term
	\begin{equation*}
	H_I=J(c_b^\dagger c_c+c_bc_c^\dagger+\gamma(2c_b^\dagger c_b-\mathbbm{1})(2c_c^\dagger c_c-\mathbbm{1})).
	\end{equation*}
    Here, $\kappa_b$ and $\kappa_c$ denote the energy gaps for the battery and charger components, respectively. $J$ denotes the strength of the interaction between the battery and the charger, while $c_{b(c)}^\dagger$ and $c_{b(c)}$ represent the creation and annihilation operators on the battery (charger) subsystem, respectively. $\gamma$ is the anisotropy parameter. It is worth noting that, in the two-excitation subspace $\{|ge\rangle,|eg\rangle\}$, the anisotropy term gives the same eigenvalue $-1$ for both states. Hence it contributes only a state-independent energy shift $-J\gamma$, which amounts to a global phase during time evolution and does not affect any observable quantities. Therefore, we set $\gamma=0.5$ in the numerical calculations without loss of generality. The term $c_b^\dagger c_c+c_bc_c^\dagger$ governs the coherent excitation transfer between the battery and the charger, and $(2c_b^\dagger c_b-\mathbbm{1})(2c_c^\dagger c_c-\mathbbm{1})$ accounts for the occupation-number coupling between the two subsystems. $f(t)$ is the switch function, defined as:
   $$
   f(t)=\left\{\begin{array}{rcl}
   	1,& t\in[0,t_0]\\
   	\\
   	0,& \text{others}~~~
   \end{array}\right.
   $$
   where $t_0$ is the time at which the interaction between the battery and the charger is terminated.
    Quantum batteries generally exhibit enhanced performance under resonant conditions \cite{lpwb,sqll}. Therefore, in this work, we focus on the resonant case where $\kappa_b=\kappa_c=\kappa$.
    
    In the initial time $t=0$, the battery is in the ground state $\rho_b(0)=|g\rangle_b$, and the charger is in the excited state $\rho_c(0)=|e\rangle_c$. Since the interaction term $H_I$ in the Hamiltonian conserves the total number of excitations, we can constrain the dynamics of the battery system to a two-dimensional subspace to simplify calculations, thereby obtaining the analytical form of the dynamically evolved state:
    \begin{equation*}
    	\begin{split}
    		|\psi(t)\rangle&=\exp(-iHt)|ge\rangle=\frac{1}{2}(e^{-i(\kappa-J-J\gamma)t}+e^{-i(\kappa+J-J\gamma)t})|ge\rangle\\
    		&+\frac{1}{2}(e^{-i(\kappa+J-J\gamma)t}-e^{-i(\kappa-J-J\gamma)t})|eg\rangle\\
    		&=\alpha(t)|ge\rangle+\beta(t)|eg\rangle.
    	\end{split}
    \end{equation*}
   
    At time $t_0$, the interaction between the battery and the charger is abruptly switched off. At this moment, two scenarios are worthy of consideration: one where the battery's acceleration process is initiated while the charger remains stationary, and the other where the battery remains stationary while the charger's acceleration process is initiated. We first consider the former.
    
    \subsection{Scenario 1}
    The accelerated battery is treated as a pointlike Unruh–DeWitt detector moving along a prescribed trajectory, with its worldline given by:
    \begin{equation}
    \begin{split}
    t(\tau)&=\frac{\sinh a\tau}{a},\,x(\tau)=\frac{\cosh a\tau}{a},\\
    y(\tau)&=z(\tau)=0,
    \end{split}
    \end{equation}
    where $\tau$ and $a$ are the battery's proper time and proper acceleration, respectively \cite{lcca,qltl,shss}. Throughout this paper, we use natural units by setting the speed of light $c$, the Planck constant $\hbar$, and the Boltzmann constant $k_B$ to unity, i.e., $c=\hbar=k_B=1$. The battery is coupled to a massless scalar quantum field $\Phi(x)$. The interaction Hamiltonian is given by:
    \begin{equation}
    H_\text{int}^b(\tau)=\varsigma(\tau)[\chi(x)c_be^{-i\kappa\tau}+\text{h.c.}]\Phi(x(\tau)).
    \end{equation}
    The switching function $\varsigma(\tau)$ is defined such that $\varsigma(\tau)=\varsigma_0$ for $t_0\leq\tau\leq t_0+\Delta$ and is zero elsewhere. The operator $c_b$ is the annihilation operator for the battery, satisfying $c_b|e\rangle_b=|g\rangle_b,\,c_b|g\rangle_b=c_b^\dagger|e\rangle_b=0,\,c_b^\dagger|g\rangle_b=|e\rangle_b$. $\chi(x)$ is the spatial coupling function which vanishes outside a small volume around the detector. Such a Gaussian coupling function describes a pointlike detector which only interacts with the neighbor scalar fields in Minkowski vacuum. During the acceleration phase, the charger remains stationary, is uncoupled from any field, and its detector is switched off. Hence, the total Hamiltonian of the system during this phase is
    \begin{equation}
    H=H_b+H_c+H_\Phi+H_\text{int}^b,
    \end{equation}
    where $H_\Phi$ is the free Hamiltonian of the scalar field.
    
    We take the state at the beginning of acceleration as the initial state, i.e.,
    \begin{equation}
    |\psi_{-\infty}^{bc\Phi}\rangle=|\psi(t_0)\rangle\otimes|0_M\rangle,
    \end{equation}
    where $|0_M\rangle$ represents the external scalar field in Minkowski vacuum. Under the weak-coupling regime, we employ a first-order perturbation in the coupling constant to compute the dynamical evolution governed by the Hamiltonian \cite{shss,wgur}, so that the final state can be written as
    \begin{equation*}
    \begin{split}
    |\psi_{\infty}^{bc\Phi}\rangle&=|\psi_{-\infty}^{bc\Phi}\rangle-i\int d\tau H_\text{int}^b(\tau)|\psi_{-\infty}^{bc\Phi}\rangle\\
    &=\alpha(t_0)|ge0_M\rangle+\beta(t_0)|eg0_M\rangle\\
    &-i\alpha(t_0)\int d\tau H_\text{int}^b(\tau)|ge0_M\rangle\\
    &-i\beta(t_0)\int d\tau H_\text{int}^b(\tau)|eg0_M\rangle.
    \end{split}
    \end{equation*}
    Note that
    \begin{equation*}
    \begin{split}
    H_\text{int}^b(\tau)|ge0_M\rangle&=\varsigma(\tau)e^{i\kappa\tau}\chi(x)\Phi(x(\tau))|ee0_M\rangle,\\
    H_\text{int}^b(\tau)|eg0_M\rangle&=\varsigma(\tau)e^{-i\kappa\tau}\chi(x)\Phi(x(\tau))|gg0_M\rangle.
    \end{split}
    \end{equation*}
    Therefore, one has
    \begin{equation*}
    	\begin{split}
    		|\psi_{\infty}^{bc\Phi}\rangle&=\alpha(t_0)|ge0_M\rangle+\beta(t_0)|eg0_M\rangle\\
    		&-i\alpha(t_0)\int d\tau \varsigma(\tau)e^{i\kappa\tau}\chi(x)\Phi(x(\tau))|ee0_M\rangle\\
    		&-i\beta(t_0)\int d\tau \varsigma(\tau)e^{-i\kappa\tau}\chi(x)\Phi(x(\tau))|gg0_M\rangle.
    	\end{split}
    \end{equation*}
    The field operators are defined as
    \begin{equation*}
    \begin{split}
    \phi(f_1)&=-i\int d\tau \varsigma(\tau)e^{i\kappa\tau}\chi(x)\Phi(x(\tau)),\\
    \phi(f_2)&=-i\int d\tau \varsigma(\tau)e^{-i\kappa\tau}\chi(x)\Phi(x(\tau)).
    \end{split}
    \end{equation*}
    Then the final state can be rewritten as
    \begin{equation}
    \begin{split}
    |\psi_{\infty}^{bc\Phi}\rangle&=\alpha(t_0)|ge0_M\rangle+\beta(t_0)|eg0_M\rangle\\
    &+\alpha(t_0)\phi(f_1)|ee0_M\rangle+\beta(t_0)\phi(f_2)|gg0_M\rangle.
    \end{split}
    \end{equation}
    
    The excited state generated by the action of the field operator on the Minkowski vacuum can be expressed as a Rindler single-particle state \cite{agsl}. Via the Bogoliubov transformation and by introducing the acceleration parameter $q=e^{\frac{-2\pi\kappa}{a}}$ and the effective coupling strength $\nu^2=\frac{\varsigma_0^2\kappa\Delta}{2\pi}e^{-\kappa^2\varkappa^2}$, we have
    \begin{equation*}
    \begin{split}
    \phi(f_1)|0_M\rangle&=\frac{\nu}{\sqrt{1-q}}|1_{F_1}\rangle,\\
    \phi(f_2)|0_M\rangle&=\frac{\nu\sqrt{q}}{\sqrt{1-q}}|1_{F_2}\rangle,
    \end{split}
    \end{equation*}
    where $\varkappa$ is the characteristic length of the coupling, representing the effective spatial size of the detector, and $|1_{F_1}\rangle$ and $|1_{F_2}\rangle$ are orthonormal single-particle states. Therefore, Eq. (7) can be written as
    \begin{equation}
    \begin{split}
    |\psi_\infty^{bc\Phi}\rangle&=\alpha(t_0)|ge0_M\rangle+\beta(t_0)|eg0_M\rangle\\
    &+\frac{\alpha(t_0)\nu}{\sqrt{1-q}}|ee1_{F_1}\rangle+\frac{\beta(t_0)\nu\sqrt{q}}{\sqrt{1-q}}|gg1_{F_2}\rangle.
    \end{split}
    \end{equation}
    By tracing out the field degrees of freedom, we obtain the reduced density matrix for the composite battery-charger system, i.e.,
    \begin{small}
    \begin{equation*}
    	\rho_{bc}=\frac{1}{N}\left(
    	\begin{array}{cccc}
    		\frac{\nu^2q|\beta(t_0)|^2}{1-q} & 0 & 0 & 0\\
    		0 & |\alpha(t_0)|^2 & \alpha(t_0)\beta(t_0)^* & 0\\
    		0 & \alpha(t_0)^*\beta(t_0) & |\beta(t_0)|^2 & 0\\
    		0 & 0 & 0 & \frac{\nu^2|\alpha(t_0)|^2}{1-q}\\
    	\end{array}
    	\right ),
    \end{equation*}
    \end{small}
    where $N=1+\frac{\nu^2}{1-q}(|\alpha(t_0)|^2+q|\beta(t_0)|^2)$ is the normalization factor. Obviously, $q$ varies monotonically with the acceleration $a$: $q\rightarrow0$ corresponds to zero acceleration, while $q\rightarrow1$ corresponds to infinite acceleration. We treat the detector–field interaction using first-order perturbation theory. To ensure the reliability of our results, we adopt the most conservative validity criterion for perturbation theory, namely, that the expansion parameter satisfies:
    \begin{equation}
    \frac{\nu^2}{1-q}\leq0.1.
    \end{equation}
    Based on this, we take $\nu^2=0.06$ for our main numerical results, for which the effective range is $q\le0.4$. Within this range, all results are obtained in the regime where perturbation theory is strictly valid.  Numerical results beyond these ranges are provided only as qualitative references for trend extrapolation and do not constitute the core evidence of our paper. Additionally, the order of basis states with respect to the density matrix we adopt is $\{|gg\rangle,\,|ge\rangle,\,|eg\rangle,\,|ee\rangle\}.$

     Now, we consider the populations of the reduced density matrix $\rho_b$ of the battery subsystem. It is straightforward to verify that the ground-state population $P_g$ and the excited-state population $P_e$ can be written as:
    	\begin{equation*}
    		\begin{split}
    			&P_g=\frac{1}{N}(|\alpha(t_0)|^2+\frac{\nu^2q|\beta(t_0)|^2}{1-q}),\\
    			&P_e=\frac{1}{N}(|\beta(t_0)|^2+\frac{\nu^2|\alpha(t_0)|^2}{1-q}).
    		\end{split}
    	\end{equation*}
    	According to the definition of ergotropy, for a two-level system, nonzero ergotropy exists only when the excited-state population exceeds the ground-state population, i.e., $P_e>P_g$. Next, we compute the difference between $P_e$ and $P_g$:
    	\begin{equation*}
    		P_e-P_g=\frac{|\beta(t_0)|^2-|\alpha(t_0)|^2+\frac{\nu^2}{1-q}(|\alpha(t_0)|^2-q|\beta(t_0)|^2)}{N}.
    	\end{equation*}
    	Let $\alpha=|\alpha(t_0)|^2$, so $|\beta(t_0)|^2=1-\alpha$. Substituting into the above and dropping the positive denominator, the critical condition $P_g=P_e$ implies:
    	\begin{equation*}
    		1-2\alpha+\frac{\nu^2}{1-q}[(1+q)\alpha-q]=0.
    	\end{equation*}
    	Based on this, we can solve for the critical acceleration parameter $q_c$, i.e., the threshold at which ergotropy suddenly emerges:
    	\begin{equation}
    		q_c=\frac{2\alpha-1-\nu^2\alpha}{\nu^2(\alpha-1)+2\alpha-1},
    	\end{equation}
    	where $\alpha=|\alpha(t_0)|^2$. This is the exact general formula, valid for any finite $t_0$. Eq. (10) implies that, for a given $\nu^2$, $q_c$ is a functional of $t_0$. By constraining $q_c$	to lie within the strictly valid perturbative regime, we can invert to solve for $t_0$, thereby identifying the critical point at which ergotropy may undergo a significant change. By restricting $q_c$	to the interval $[0,0.4]$, we numerically solve for the corresponding range of $t_0$ within the first period as $t_0\in(0.76927,0.76994)$.
    	
    	\begin{figure}[htbp]
    		\centering
    		\includegraphics[width=0.5\textwidth]{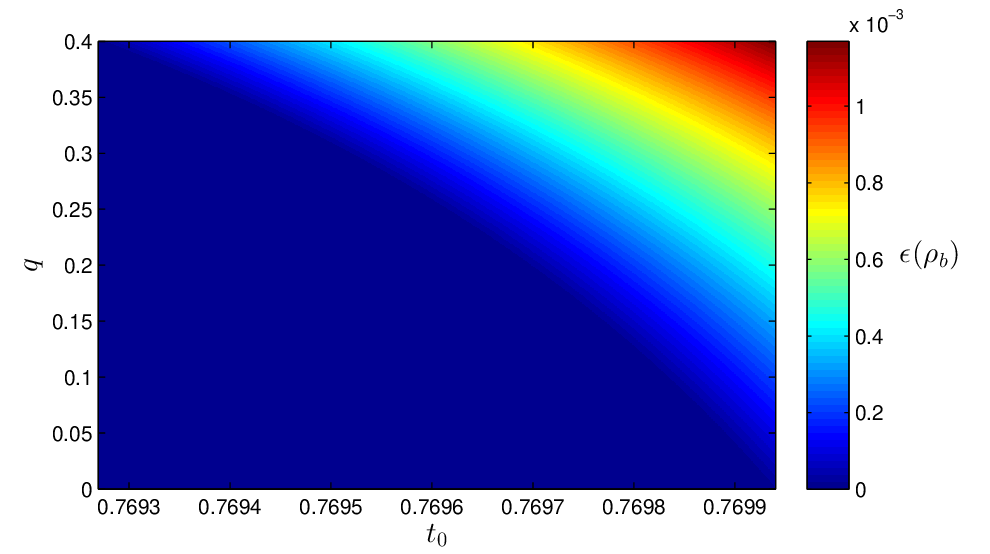}
    		\vspace{-2em} \caption{Battery subsystem ergotropy $\epsilon(\rho_b)$ (in units of $\kappa$) as a function of the acceleration onset time $t_0$ and the acceleration parameter $q$, evaluated with respect to the battery local Hamiltonian $h_b$. The effective coupling strength is $\nu^2=0.06$, with $\kappa=J=1$, $\gamma=0.5$. The acceleration parameter $q=0.4$, corresponding to $\frac{\nu^2}{1-q}=0.1$, the boundary of the strictly valid perturbative regime.} \label{Fig.1}
    	\end{figure}
    	Figure 1 plots the battery subsystem ergotropy as a function of the acceleration parameter $q$ and $t_0$, with $q\in[0,0.4]$ and $t_0\in(0.76927,0.76994)$. The battery subsystem is described by the Hamiltonian $H_b$. We find that the ergotropy of the battery subsystem suddenly emerges when $q$ exceeds a certain critical value. The Unruh effect causes the accelerated detector to experience a thermal bath, pumping population from the ground state to the excited state. When the acceleration exceeds the critical value, it induces population inversion. Once the inversion threshold is crossed, the passive state becomes active, and ergotropy suddenly emerges.

   To further understand the physical mechanism of the Unruh effect on the energy of the composite system, we compute the total energy change of the composite system. Tracing over the field degrees of freedom in Eq. (8) yields the reduced density matrix of the composite system, and thus the expectation value of its energy:
   \begin{equation}
   E_{bc}(\rho)=\kappa\frac{1+\frac{2\nu^2|\alpha(t_0)|^2}{1-q}}{1+\frac{\nu^2}{1-q}(|\alpha(t_0)|^2+q|\beta(t_0)|^2)}.
   \end{equation}
   Here, with the interaction turned off, we compute the energy using only the composite-system Hamiltonian $H_b+H_c$. At time $t_0$, acceleration has not yet begun and the detector is decoupled from the field. The system is in the pure state $|\psi(t_0)\rangle=\alpha|ge\rangle+\beta|eg\rangle$, so the energy at this instant can be written as:
   \begin{equation}
   E_{bc}(t_0)=\kappa(|\alpha(t_0)|^2+|\beta(t_0)|^2)=\kappa.
   \end{equation}
   Therefore, the energy change induced by the acceleration phase is
   \begin{equation*}
   	\begin{split}
   	\Delta E_{bc}(\rho)&=E_{bc}(\rho)-E_{bc}(t_0)\\
   	&=\kappa(\frac{1+\frac{2\nu^2|\alpha(t_0)|^2}{1-q}}{1+\frac{\nu^2}{1-q}(|\alpha(t_0)|^2+q|\beta(t_0)|^2)}-1)\\
   	&=\kappa\frac{\nu^2(|\alpha(t_0)|^2-q|\beta(t_0)|^2)}{1-q+\nu^2(|\alpha(t_0)|^2+q|\beta(t_0)|^2)}.
   	\end{split}
   \end{equation*}
   
   Figure 2 shows the composite-system energy change $\Delta E_{bc}(\rho)$ as a function of the acceleration parameter $q$ for two values of $t_0$: $\frac{\pi}{8}$ and $\frac{3\pi}{8}$. First, for $t_0=\pi/8$, $\Delta E_{bc}(\rho)$ is already positive at $q=0$ and increases monotonically with $q$. This indicates that even without acceleration, the switching of the detector–field coupling has already injected a small amount of energy into the composite system via vacuum fluctuations. As acceleration increases, the Unruh effect further amplifies this energy injection, resulting in a net energy gain throughout the acceleration phase that grows steadily with $q$. Second, for $t_0=\frac{3\pi}{8}$, $\Delta E_{bc}$ exhibits a markedly different behavior: at small $q$, the composite system still achieves a net energy gain, but as $q$ increases, $\Delta E_{bc}$ decreases monotonically. Once $q$ exceeds a certain threshold, $\Delta E_{bc}$ changes from positive to negative, and the composite system begins to experience a net energy loss.   
    \begin{figure}[htbp]
    	\centering
    	\includegraphics[width=0.5\textwidth]{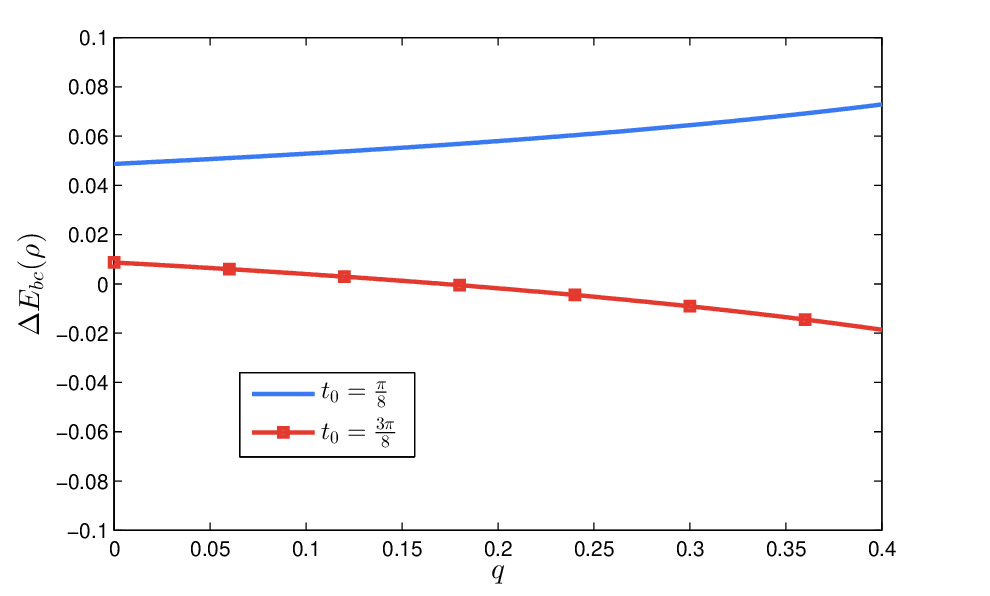}
    	\vspace{-2em} \caption{Composite-system stored energy $\Delta E_{bc}(\rho)$ (in units of $\kappa$) as a function of the acceleration parameter $q$ for two values of $t_0$: $\frac{\pi}{8}$ and $\frac{3\pi}{8}$. The energy is evaluated with respect to the post-switch-off Hamiltonian $H_b+H_c$, with the initial energy at $t_0$ subtracted as the reference. Parameters: $\nu^2=0.06$, $\kappa=J=1$, $\gamma=0.5$. The acceleration $q=0.4$, the boundary of the strictly valid perturbative regime.} \label{Fig.3}
    \end{figure}
    
    The above energy change analysis implies that the increase in the composite system energy is supplied by an external driving source that maintains the detector's accelerated motion. The Unruh effect redistributes this input mechanical energy into internal excitations of the battery and charger subsystems. However, our framework only quantifies the stored energy variation with acceleration and does not account for the external work consumed by the acceleration itself. A complete thermodynamic energy budget including the cost of acceleration is left for future work.

    \subsection{Scenario 2}
    Now, we consider the scenario where the battery remains stationary and the charger's detector is switched on at time $t_0$, i.e., the charger begins to accelerate. The Hamiltonian describing the interaction between the charger and the massless scalar quantum field can be written as:
    \begin{equation}
    H_\text{int}^c(\tau)=\varsigma(\tau)[\chi(x)c_ce^{-i\kappa\tau}+\text{h.c.}]\Phi(x(\tau)),
    \end{equation}
    where $c_c$ is the annihilation operator for the charger. The initial state at the onset of accelerated motion is still given by Eq. (6). We adopt the same first-order perturbation method as in Scenario 1, which yields the final state in the form:
     \begin{equation}
    	\begin{split}
    		|\psi_\infty^{bc\Phi}\rangle&=\alpha(t_0)|ge0_M\rangle+\beta(t_0)|eg0_M\rangle\\
    		&+\frac{\alpha(t_0)\nu\sqrt{q}}{\sqrt{1-q}}|gg1_{F_2}\rangle+\frac{\beta(t_0)\nu}{\sqrt{1-q}}|ee1_{F_1}\rangle.
    	\end{split}
    \end{equation}
    By tracing out the field degrees of freedom, we obtain the reduced density matrix for the composite battery-charger system, i.e.,
    \begin{small}
    	\begin{equation*}
    		\rho_{bc}=\frac{1}{N}\left(
    		\begin{array}{cccc}
    			\frac{\nu^2q|\alpha(t_0)|^2}{1-q} & 0 & 0 & 0\\
    			0 & |\alpha(t_0)|^2 & \alpha(t_0)\beta(t_0)^* & 0\\
    			0 & \alpha(t_0)^*\beta(t_0) & |\beta(t_0)|^2 & 0\\
    			0 & 0 & 0 & \frac{\nu^2|\beta(t_0)|^2}{1-q}\\
    		\end{array}
    		\right ).
    	\end{equation*}
    \end{small}
    For simplicity, the normalization factor is still denoted by $N$, namely, $N=1+\frac{\nu^2}{1-q}(q|\alpha(t_0)|^2+|\beta(t_0)|^2)$.
    
    Next, we examine the population distribution of the battery subsystem. Tracing over the charger and field degrees of freedom in Eq. (16) yields the difference between the excited-state and ground-state populations of the battery:
    \begin{equation*}
    P_e-P_g=\frac{2|\beta(t_0)|^2-1+\frac{\nu^2}{1-q}[(1+q)|\beta(t_0)|^2-q]}{1+\frac{\nu^2}{1-q}(q|\alpha(t_0)|^2+|\beta(t_0)|^2)}.
    \end{equation*}
    Within the perturbatively valid regime, we expand the above expression to order $\nu^2$. Defining $\eta=\frac{\nu^2}{1-q}$ and using 
    \begin{small}
    $$
    \frac{1}{1+\eta(q|\alpha(t_0)|^2+|\beta(t_0)|^2)}=1-\eta(q|\alpha(t_0)|^2+|\beta(t_0)|^2)+O(\nu^2),
    $$ 
    \end{small}
    we obtain:
    \begin{small}
    \begin{equation}
    	\begin{split}
    		P_e-P_g&=2|\beta(t_0)|^2-1+2\eta(1-q)|\alpha(t_0)|^2|\beta(t_0)|^2+O(\eta^2)\\
    		&=2|\beta(t_0)|^2-1+2\nu^2|\alpha(t_0)|^2|\beta(t_0)|^2+O(\nu^4).
    	\end{split}
    \end{equation}
    \end{small}
    Since our perturbation theory strictly restricts $q$ to be far from 1 ($q<0.4$), the second equality is guaranteed. Note that the result in Eq. (15) is of great importance: the effect of the acceleration parameter $q$ on the battery population difference is suppressed to order $O(\nu^4)$ and is completely indistinguishable within the effective accuracy of perturbation theory. Hence:
    \\
     (i) if the battery is initially in a passive state, then $P_e-P_g\le0$ for all $q$, so the battery remains passive throughout the acceleration and its ergotropy is always zero; 
     \\
     (ii) if the battery is initially in an active state, its maximum extractable work (ergotropy) remains constant and does not change with $q$.
     \\
     This analytical result rigorously proves that the accelerated charger cannot induce population inversion in the battery. Consequently, in Scenario 2, the ergotropy of the battery subsystem remains constant with respect to $q$, and no sudden emergence phenomenon such as that observed in Scenario 1 occurs.

    \subsection{Scenario 3}
    Now consider the third scenario: at time $t_0$, both the battery and the charger begin uniform acceleration simultaneously, and both are coupled to a massless scalar quantum field. Their worldlines are identical and given by:
    \begin{equation*}
    	\begin{split}
    		t(\tau)&=\frac{\sinh a\tau}{a},\,x(\tau)=\frac{\cosh a\tau}{a},\\
    		y(\tau)&=z(\tau)=0,
    	\end{split}
    \end{equation*}
    where $\tau$ and $a$ are the proper time and proper acceleration, respectively. The two detectors are simultaneously coupled to the field, and the total interaction Hamiltonian is given by:
    \begin{equation}
    	H_\text{int}^{bc}(\tau)=\varsigma(\tau)[\chi(x)(c_b+c_c)e^{-i\kappa\tau}+\text{h.c.}]\Phi(x(\tau)).
    \end{equation}
    The initial state at the onset of the acceleration phase is:
    \begin{equation*}
    |\psi_{-\infty}^{bc\Phi}\rangle=|\psi(t_0)\rangle\otimes|0_M\rangle.
    \end{equation*}
    Under weak coupling, using first-order perturbation theory, the final state is:
    \begin{equation*}
    \begin{split}
    |\psi_{\infty}^{bc\Phi}\rangle&=|\psi_{-\infty}^{bc\Phi}\rangle-i\int d\tau H_\text{int}^{bc}(\tau)|\psi_{-\infty}^{bc\Phi}\rangle\\
    &=\alpha(t_0)|ge0_M\rangle+\beta(t_0)|eg0_M\rangle\\
    &-i\alpha(t_0)\int d\tau H_\text{int}^{b}(\tau)|ge0_M\rangle-i\beta(t_0)\int d\tau H_\text{int}^{b}(\tau)|eg0_M\rangle\\
    &-i\alpha(t_0)\int d\tau H_\text{int}^{c}(\tau)|ge0_M\rangle-i\beta(t_0)\int d\tau H_\text{int}^{c}(\tau)|eg0_M\rangle\\
    \end{split}
    \end{equation*}
    Computing each term separately, we obtain that
    \begin{equation*}
    \begin{split}
    H_\text{int}^{b}(\tau)|ge0_M\rangle&=\varsigma(\tau)e^{i\kappa\tau}\chi(x)\Phi(x(\tau))|ee0_M\rangle,\\
    H_\text{int}^{b}(\tau)|eg0_M\rangle&=\varsigma(\tau)e^{-i\kappa\tau}\chi(x)\Phi(x(\tau))|gg0_M\rangle,\\
    H_\text{int}^{c}(\tau)|ge0_M\rangle&=\varsigma(\tau)e^{-i\kappa\tau}\chi(x)\Phi(x(\tau))|gg0_M\rangle,\\
    H_\text{int}^{c}(\tau)|eg0_M\rangle&=\varsigma(\tau)e^{i\kappa\tau}\chi(x)\Phi(x(\tau))|ee0_M\rangle.
    \end{split}
    \end{equation*}
    Thus the final state can be rewritten as
    \begin{equation*}
    \begin{split}
    |\psi_{\infty}^{bc\Phi}\rangle&=\alpha(t_0)|ge0_M\rangle+\beta(t_0)|eg0_M\rangle\\
    &+\alpha(t_0)\phi(f_1)|ee0_M\rangle+\beta(t_0)\phi(f_2)|gg0_M\rangle\\
    &+\alpha(t_0)\phi(f_2)|gg0_M\rangle+\beta(t_0)\phi(f_1)|ee0_M\rangle\\
    &=\alpha(t_0)|ge0_M\rangle+\beta(t_0)|eg0_M\rangle\\
    &+\frac{\nu}{\sqrt{1-q}}(\alpha(t_0)+\beta(t_0))|ee1_{F_1}\rangle\\
    &+\frac{\nu\sqrt{q}}{\sqrt{1-q}}(\alpha(t_0)+\beta(t_0))|gg1_{F_2}\rangle,
    \end{split}
    \end{equation*}
    where in the second equality we have used the Bogoliubov transformation. Tracing over the field degrees of freedom, we obtain the reduced density matrix of the composite system. The basis order is $\{|gg\rangle, |ge\rangle, |eg\rangle, |ee\rangle\}$:
    \begin{small}
    	\begin{equation*}
    		\rho_{bc}=\frac{1}{N}\left(
    		\begin{array}{cccc}
    			\frac{\nu^2q}{1-q} & 0 & 0 & 0\\
    			0 & |\alpha(t_0)|^2 & \alpha(t_0)\beta(t_0)^* & 0\\
    			0 & \alpha(t_0)^*\beta(t_0) & |\beta(t_0)|^2 & 0\\
    			0 & 0 & 0 & \frac{\nu^2}{1-q}\\
    		\end{array}
    		\right ),
    	\end{equation*}
    \end{small}
    where $N=1+\frac{\nu^2}{1-q}(1+q)$ is the normalization factor. 
    
    We now analyze the emergence conditions for ergotropy in Scenario 3. To this end, we first compute the population difference of the battery subsystem:
    \begin{equation*}
    \begin{split}
    P_e-P_g&=\frac{|\beta(t_0)|^2-|\alpha(t_0)|^2+\frac{\nu^2}{1-q}(1-q)}{1+\frac{\nu^2}{1-q}(1+q)}\\
    &=\frac{|\beta(t_0)|^2-|\alpha(t_0)|^2+\nu^2}{1+\frac{\nu^2}{1-q}(1+q)}\\
    &=\frac{2|\beta(t_0)|^2-1+\nu^2}{1+\frac{\nu^2}{1-q}(1+q)}.
    \end{split}
    \end{equation*}
    The critical condition for ergotropy emergence is $P_e-P_g=0$. Since the denominator is always positive, $P_e-P_g=0$ implies $|\beta(t_0)|^2=\frac{1-\nu^2}{2}$. Note that this result reveals the essential difference between Scenario 3 and Scenario 1. In Scenario 1, for a fixed population $|\beta(t_0)|$, ergotropy may suddenly emerge at a finite critical acceleration $q_c$. In Scenario 3, however, no such $q_c$ exists. If $|\beta(t_0)|>\frac{1-\nu^2}{2}$, ergotropy is nonzero for all $q\in[0,0.4]$; if $|\beta(t_0)|<\frac{1-\nu^2}{2}$, it is zero for all $q$. In other words, the presence or absence of ergotropy is determined entirely by the initial population at the onset of acceleration and is independent of the acceleration parameter $q$.
    
    Next, we analyze the dynamical behavior of the battery subsystem ergotropy as a function of the acceleration parameter $q$ in the co-acceleration scenario. From the above derivation, the population difference of the battery subsystem is given by:
    \begin{equation}
    P_e-P_g=\frac{2|\beta(t_0)|^2-1+\nu^2}{1+\frac{\nu^2}{1-q}(1+q)}.
    \end{equation}
    Note that, for fixed $t_0$ and $\nu^2$, the numerator on the right-hand side is a constant with respect to the acceleration parameter $q$. For brevity, we denote $C=2|\beta(t_0)|^2-1+\nu^2$. Then, one has
    \begin{equation}
    P_e-P_g=\frac{C}{1+\frac{\nu^2}{1-q}(1+q)}.
    \end{equation}
    We differentiate the denominator with respect to $q$:
    \begin{equation*}
    (1+\frac{\nu^2}{1-q}(1+q))^{'}=\frac{2\nu^2}{(1-q)^2}>0.
    \end{equation*}
    Since $C$ is independent of $q$, we further have
    \begin{equation}
    \frac{d}{dq}(P_e-P_g)=-\frac{C.\frac{2\nu^2}{(1-q)^2}}{(1+\frac{\nu^2}{1-q}(1+q))^{2}}.
    \end{equation}
    It is straightforward to verify that the battery subsystem ergotropy can be rewritten as:
    \begin{equation}
    \epsilon(\rho_b)=\kappa.\max\{P_e-P_g,0\}.
    \end{equation}
    Combining Eqs. (18), (19), and (20), we find that for $C>0$, the ergotropy $\epsilon(\rho_b)$ is strictly monotonically decreasing with $q$, while for $C\leq0$, $\epsilon(\rho_b)=0$ for all $q$. This analytical result is consistent with the behavior shown in Figure 3.
    \begin{figure}[htbp]
    	\centering
    	\includegraphics[width=0.5\textwidth]{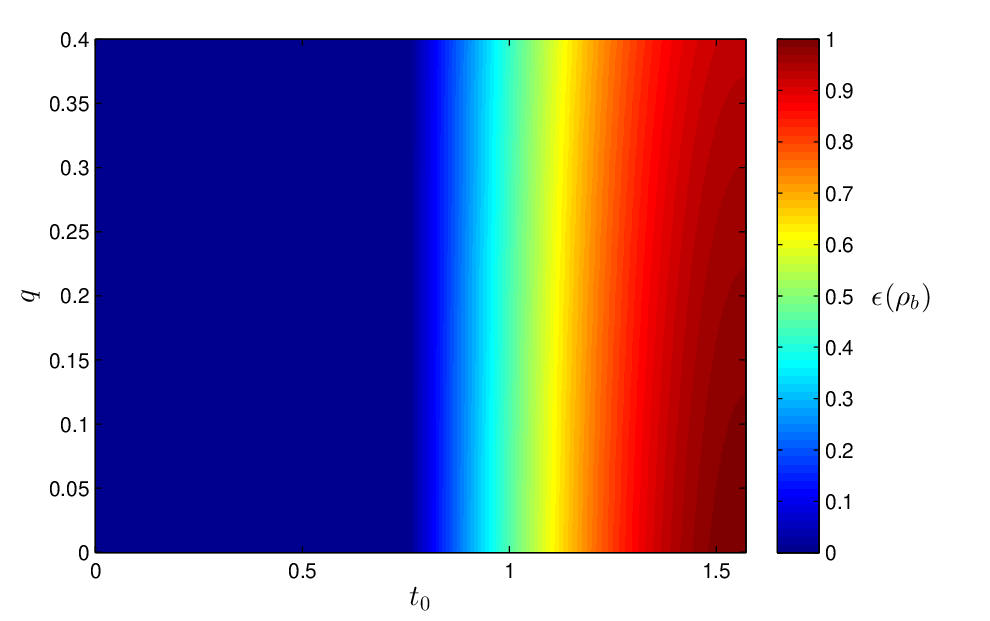}
    	\vspace{-2em} \caption{Battery subsystem ergotropy $\epsilon(\rho_b)$ (in units of $\kappa$) as a function of $t_0$ and the acceleration parameter $q$ for the bipartite simultaneous acceleration scenario (Scenario 3), evaluated with respect to the battery local Hamiltonian $H_b$. Parameters: $\nu^2=0.06$, $\kappa=J=1$, $\gamma=0.5$. The acceleration parameter $q=0.4$, the boundary of the strictly valid perturbative regime.} \label{Fig.2}
    \end{figure}
    
    In Scenario 3, the battery and charger are both coupled to the scalar field. In the unnormalized total state, the coefficient of the field-excited state $|ee1_{F_1}\rangle$ is $\frac{\nu(\alpha+\beta)}{\sqrt{1-q}}$, arising from the coherent superposition of transition amplitudes from the two initial components $|ge\rangle$ and $|eg\rangle$ to the same final state. Since $|\alpha+\beta|^2=1$, this coherent superposition exactly cancels all $q$-dependent terms in the numerator after normalization, which is equivalent to a uniform dilution of the system by a global thermal bath. The sole effect of the acceleration parameter $q$ is to increase the denominator, i.e., to redistribute probability weights from the lower-energy components to the field-excited components during normalization. This redistribution produces a proportional dilution of the population polarization. When the numerator is positive, this dilution causes the ergotropy to decrease monotonically with $q$; when the numerator is nonpositive, the ergotropy is identically zero.
    
    It is worth noting that the difference between global and local ergotropy, $\Delta\epsilon_\text{corr}=\epsilon(\rho_{bc})-\epsilon(\rho_b)-\epsilon(\rho_c)$, is an interesting direction. However, for the density matrix arising from the Unruh effect, there is no simple analytical relation between $\Delta\epsilon_\text{corr}$ and standard entanglement measures such as concurrence. In the future, connecting quantum correlations such as nonlocality to ergotropy \cite{gf,rpas,atbc,bvrs} may offer new insights into the relationship between quantum correlations and work extraction under the Unruh effect.
    \section{III. Unruh Effect on Tripartite Quantum Batteries}
    In this section, we investigate the Unruh effect on a tripartite quantum battery system. The Hamiltonian $H$ consists of a two-level system representing the battery and two two-level systems representing the chargers, which interact via nearest-neighbor Heisenberg XX-type coupling:
    \begin{equation*}
    	\begin{split}
    		H&=H_a+H_b+H_c+f(t)H_I\\
    		&=\kappa_ac_a^\dagger c_a+\kappa_bc_b^\dagger c_b+\kappa_cc_c^\dagger c_c+f(t)H_I,
    	\end{split}
    \end{equation*}
    with the coupling term
    \begin{equation*}
    	\begin{split}
    		H_I&=J_1(c_a^\dagger c_b+c_ac_b^\dagger)+J_2(c_b^\dagger c_c+c_bc_c^\dagger).
    	\end{split}
    \end{equation*}
    Note that, different from the bipartite case, here $H_a$ represents the free Hamiltonian of the battery subsystem, while $H_b$ and $H_c$ are taken to be the free Hamiltonians of the two chargers. $\kappa_a$, $\kappa_b$, and $\kappa_c$ represent the energy gaps for the battery and two charger parts, respectively. $J_1$ denotes the coupling strength between the battery and neighboring charger, and $J_2$ represents the strength between the two chargers. $f(t)$ is the switch function, consistent with its earlier definition. We still focus on the resonant condition, i.e., $\kappa_a=\kappa_b=\kappa_c=\kappa$.
    
    Initially, the battery is prepared in the ground state $\varrho_a(0)=|g\rangle_a$, the neighboring charger is also in the ground state $\varrho_b(0)=|g\rangle_b$, and the another charger is  in the excited state $\varrho_c(0)=|e\rangle_c$. Therefore, the total state at the initial time is $\varrho(0)=|g\rangle_a|g\rangle_b|e\rangle_c$. The derivation details of the tripartite quantum battery system dynamics are given in Appendix A. Then, the state of the total system at time $t$ can be written as:
    \begin{small}
    	\begin{equation*}
    		\begin{split}
    			|\psi(t)\rangle_{abc}&=\exp(-iHt)|gge\rangle\\
    			&=\frac{-J_1e^{-ie_1t}}{\sqrt{J_1^2+J_2^2}}|e_1\rangle+\frac{J_2e^{-ie_2t}}{\sqrt{2(J_1^2+J_2^2)}}|e_2\rangle+\frac{J_2e^{-ie_3t}}{\sqrt{2(J_1^2+J_2^2)}}|e_3\rangle\\
    			&=\alpha_1(t)|gge\rangle+\alpha_2(t)|geg\rangle+\alpha_3(t)|egg\rangle.
    		\end{split}
    	\end{equation*}
    \end{small}
    For the order of basis states with respect to the density matrix we choose is $\{|ggg\rangle, |gge\rangle, |geg\rangle, |gee\rangle, |egg\rangle, |ege\rangle, |eeg\rangle, |eee\rangle\}$, the corresponding density matrix $\varrho_{abc}(t)=|\psi(t)\rangle_{abc}\langle\psi(t)|$ can be written as
    \begin{equation}\label{e21}
    	\varrho_{abc}(t)=\left(
    	\begin{array}{cccccccc}
    		0 & 0 & 0 & 0 & 0 & 0 & 0 & 0\\
    		0 & \varrho_{11} & \varrho_{12} & 0 & \varrho_{14} & 0 & 0 & 0\\
    		0 & \varrho_{21} & \varrho_{22} & 0 & \varrho_{24} & 0 & 0 & 0\\
    		0 & 0 & 0 & 0 & 0 & 0 & 0 & 0\\
    		0 & \varrho_{41} & \varrho_{42} & 0 & \varrho_{44} & 0 & 0 & 0\\
    		0 & 0 & 0 & 0 & 0 & 0 & 0 & 0\\
    		0 & 0 & 0 & 0 & 0 & 0 & 0 & 0\\
    		0 & 0 & 0 & 0 & 0 & 0 & 0 & 0\\
    	\end{array}
    	\right ),
    \end{equation}
    where
    \begin{small}
    	\begin{equation*}
    		\begin{split}
    			\varrho_{11}&=\frac{4J_1^4+2J_2^4+8J_1^2J_2^2\cos\sqrt{J_1^2+J_2^2}t+2J_2^4\cos2\sqrt{J_1^2+J_2^2}t}{4(J_1^2+J_2^2)^2},\\
    			\varrho_{12}&=\frac{i4J_1^2J_2\sin\sqrt{J_1^2+J_2^2}t+i2J_2^3\sin2\sqrt{J_1^2+J_2^2}t}{4(J_1^2+J_2^2)^{\frac{3}{2}}},\\
    			\varrho_{14}&=\frac{J_1J_2}{4(J_1^2+J_2^2)^2}(-4J_1^2+2J_2^2+4(J_1^2-J_2^2)\cos\sqrt{J_1^2+J_2^2}t\\
    			&+2J_2^4\cos2\sqrt{J_1^2+J_2^2}t),\\
    			\varrho_{22}&=\frac{2J_2^2-2J_2^2\cos2\sqrt{J_1^2+J_2^2}t}{4(J_1^2+J_2^2)},\\
    			\varrho_{24}&=\frac{i4J_2\sin\sqrt{J_1^2+J_2^2}t-i2J_2\sin2\sqrt{J_1^2+J_2^2}t}{4(J_1^2+J_2^2)^{\frac{3}{2}}},\\
    			\varrho_{21}&=\varrho_{12}^*,\,\varrho_{41}=\varrho_{14}^*,\,\varrho_{42}=\varrho_{24}^*,\,\varrho_{44}=1-\varrho_{11}-\varrho_{22}.
    		\end{split}
    	\end{equation*}
    \end{small}
    
    At time $t_0$, the interaction between the subsystems is switched off. According to the principle that one subsystem acceleration process is initiated while the other two remain stationary, we can distinguish three scenarios. We first investigate the first scenario: the battery subsystem begins uniform acceleration while both chargers remain stationary.
    
    The accelerated battery is coupled to a massless scalar quantum field $\Phi(x)$, with the interaction Hamiltonian given by:
    \begin{equation}
    	H_\text{int}^a(\tau)=\varsigma(\tau)[\chi(x)c_ae^{-i\kappa\tau}+\text{h.c.}]\Phi(x(\tau)),
    \end{equation}
    where $c_a$ denotes the annihilation operator of the battery. The initial state at the onset of accelerated motion is
    \begin{equation}
    	|\psi_{1,-\infty}^{abc\Phi}\rangle=|\psi(t_0)\rangle_{abc}\otimes|0_M\rangle.
    \end{equation}
    We still employ the first-order perturbation approximation to compute the final state of the system, which can be expressed as the sum of the initial state and the first-order correction term, i.e.,
    \begin{small}
    	\begin{equation*}
    		\begin{split}
    			|&\psi_{1,\infty}^{abc\Phi}\rangle=\alpha_1(t_0)|gge0_M\rangle+\alpha_2(t_0)|geg0_M\rangle+\alpha_3(t_0)|egg0_M\rangle\\
    			&+\frac{\alpha_1(t_0)\nu}{\sqrt{1-q}}|ege1_{F_1}\rangle+\frac{\alpha_2(t_0)\nu}{\sqrt{1-q}}|eeg1_{F_1}\rangle+\frac{\alpha_3(t_0)\nu\sqrt{q}}{\sqrt{1-q}}|ggg1_{F_2}\rangle.
    		\end{split}
    	\end{equation*}
    \end{small}

    By tracing out the field degrees of freedom, the density matrix of the composite system can be written as:
    \begin{equation*}
    	\varrho_{abc}^1=\frac{1}{N_1}\left(
    	\begin{array}{cccccccc}
    		\frac{\nu^2q\varrho_{44}}{1-q} & 0 & 0 & 0 & 0 & 0 & 0 & 0\\
    		0 & \varrho_{11} & \varrho_{12} & 0 & \varrho_{14} & 0 & 0 & 0\\
    		0 & \varrho_{21} & \varrho_{22} & 0 & \varrho_{24} & 0 & 0 & 0\\
    		0 & 0 & 0 & 0 & 0 & 0 & 0 & 0\\
    		0 & \varrho_{41} & \varrho_{42} & 0 & \varrho_{44} & 0 & 0 & 0\\
    		0 & 0 & 0 & 0 & 0 & \frac{\nu^2\varrho_{11}}{1-q} & \frac{\nu^2\varrho_{12}}{1-q} & 0\\
    		0 & 0 & 0 & 0 & 0 & \frac{\nu^2\varrho_{21}}{1-q} & \frac{\nu^2\varrho_{22}}{1-q} & 0\\
    		0 & 0 & 0 & 0 & 0 & 0 & 0 & 0\\
    	\end{array}
    	\right ),
    \end{equation*}
    where $N_1=1+\frac{\nu^2}{1-q}(\varrho_{11}+\varrho_{22}+q\varrho_{44})$. 
    
    For Scenario 2, where the charger adjacent to the battery initiates uniform acceleration at time $t_0$, the corresponding final state can be written as:
    \begin{small}
    	\begin{equation*}
    		\begin{split}
    			|&\psi_{2,\infty}^{abc\Phi}\rangle=\alpha_1(t_0)|gge0_M\rangle+\alpha_2(t_0)|geg0_M\rangle+\alpha_3(t_0)|egg0_M\rangle\\
    			&+\frac{\alpha_1(t_0)\nu}{\sqrt{1-q}}|gee1_{F_1}\rangle+\frac{\alpha_2(t_0)\nu\sqrt{q}}{\sqrt{1-q}}|ggg1_{F_2}\rangle+\frac{\alpha_3(t_0)\nu}{\sqrt{1-q}}|eeg1_{F_1}\rangle.
    		\end{split}
    	\end{equation*}
    \end{small}
    The corresponding density matrix of the composite system is given by:
    \begin{equation*}
    	\varrho_{abc}^2=\frac{1}{N_2}\left(
    	\begin{array}{cccccccc}
    		\frac{\nu^2q\varrho_{22}}{1-q} & 0 & 0 & 0 & 0 & 0 & 0 & 0\\
    		0 & \varrho_{11} & \varrho_{12} & 0 & \varrho_{14} & 0 & 0 & 0\\
    		0 & \varrho_{21} & \varrho_{22} & 0 & \varrho_{24} & 0 & 0 & 0\\
    		0 & 0 & 0 & \frac{\nu^2\varrho_{11}}{1-q} & 0 & 0 & \frac{\nu^2\varrho_{14}}{1-q} & 0\\
    		0 & \varrho_{41} & \varrho_{42} & 0 & \varrho_{44} & 0 & 0 & 0\\
    		0 & 0 & 0 & 0 & 0 & 0 & 0 & 0\\
    		0 & 0 & 0 & \frac{\nu^2\varrho_{41}}{1-q} & 0 & 0 & \frac{\nu^2\varrho_{44}}{1-q} & 0\\
    		0 & 0 & 0 & 0 & 0 & 0 & 0 & 0\\
    	\end{array}
    	\right ),
    \end{equation*}
    where $N_2=1+\frac{\nu^2}{1-q}(\varrho_{11}+q\varrho_{22}+\varrho_{44})$.
    
    Now, we consider the third scenario: at time $t_0$, both the battery and the charger begin uniform acceleration simultaneously, and both are coupled to a massless scalar quantum field. The two detectors are simultaneously coupled to the field, and the total interaction Hamiltonian is given by:
    \begin{equation}
    	H_\text{int}^{ab}(\tau)=\varsigma(\tau)[\chi(x)(c_a+c_b)e^{-i\kappa\tau}+\text{h.c.}]\Phi(x(\tau)).
    \end{equation}
    Thus the final state can be rewritten as
    \begin{equation*}
    	\begin{split}
    		|\psi_{\infty}^{ab\Phi}\rangle&=\alpha_1(t_0)|gge0_M\rangle+\alpha_2(t_0)|geg0_M\rangle+\alpha_3(t_0)|egg0_M\rangle\\
    		&+\frac{\nu}{\sqrt{1-q}}\alpha_1(t_0)|ege1_{F_1}\rangle+\frac{\nu}{\sqrt{1-q}}\alpha_1(t_0)|gee1_{F_1}\rangle\\
    		&+\frac{\nu}{\sqrt{1-q}}(\alpha_2(t_0)+\alpha_3(t_0))|eeg1_{F_1}\rangle\\
    		&+\frac{\nu\sqrt{q}}{\sqrt{1-q}}(\alpha_2(t_0)+\alpha_3(t_0))|ggg1_{F_2}\rangle.
    	\end{split}
    \end{equation*}

    We now investigate the Unruh effect on the ergotropy of the battery subsystem under the three scenarios. 
    \subsection{Scenario 1}
    According to the explicit form of the composite-system density matrix $\varrho_{abc}^1$ given above, the reduced density matrix of the battery subsystem can be written as:
    \begin{small}
    	\begin{equation*}
    		\rho_{a}^1=\frac{1}{N_1}\left(
    		\begin{array}{cc}
    			\varrho_{11}+\varrho_{22}+\frac{\nu^2q}{1-q}\varrho_{44} & 0\\
    			0 & \frac{\nu^2q}{1-q}(\varrho_{11}+\varrho_{22})+\varrho_{44}\\
    		\end{array}
    		\right ).
    	\end{equation*}
    \end{small}
    The population difference of the battery subsystem is:
    \begin{equation*}
    	\begin{split}
    		P_e-P_g&=\frac{\varrho_{44}-\varrho_{11}-\varrho_{22}+\frac{\nu^2}{1-q}(\varrho_{11}+\varrho_{22}-q\varrho_{44})}{1+\frac{\nu^2}{1-q}(\varrho_{11}+\varrho_{22}+q\varrho_{44})}\\
    		&=\frac{1-2X+\frac{\nu^2}{1-q}(X-q(1-X))}{1+\frac{\nu^2}{1-q}(X+q(1-X))},
    	\end{split}
    \end{equation*}
    where the second equality uses the substitution $X=\varrho_{11}+\varrho_{22}$. 
    
    We now discuss the critical condition under which ergotropy may suddenly emerge. That is to say, setting the numerator in the above expression to zero gives:
    \begin{equation*}
    	1-2X+\frac{\nu^2}{1-q_c}(X-q_c(1-X))=0.
    \end{equation*}
    This means that
    \begin{equation}
    	q_c=\frac{1-2X+\nu^2X}{1-2X+\nu^2(1-X)}.
    \end{equation}
    Note that the physical meaning of $X$ is the ground-state population of the battery subsystem before acceleration, corresponding to $|\alpha(t_0)|^2$ in Scenario 1 of the bipartite quantum battery system. Therefore, the ergotropy emergence condition in the three-body Scenario 1 is mathematically identical to that in the two-body Scenario 1. The two chargers do not participate in any dynamical processes during the acceleration phase, since they remain static and are not coupled to the field. Their only role is to influence the state of the battery at $t_0$ through the prior internal interaction. Once acceleration begins, the battery evolution is exactly the same as in the bipartite case.
    
    Furthermore, we consider the energy change of the composite system during this acceleration process. The composite-system Hamiltonian is $H_a+H_b+H_c$. The energy expectation value of the composite system can be written as:
    \begin{equation*}
    	\begin{split}
    		E_{abc}(q)&=\frac{\kappa}{N_1}[\varrho_{11}+\varrho_{22}+\varrho_{44}+\frac{2\nu^2}{1-q}(\varrho_{11}+\varrho_{22})]\\
    		&=\kappa\frac{1+\frac{2\nu^2}{1-q}X}{1+\frac{\nu^2}{1-q}(X+q(1-X))}
    	\end{split}
    \end{equation*}
    Before the acceleration process begins, the total excitation number of the composite system is 1, so the energy change during the acceleration is:
    \begin{equation*}
    	\begin{split}
    		\Delta E_{abc}(q)&=E_{abc}(q)-E_{abc}(t_0)\\
    		&=\kappa[\frac{1+\frac{2\nu^2}{1-q}X}{1+\frac{\nu^2}{1-q}(X+q(1-X))}-1]\\
    		&=\kappa\frac{\nu^2[X-q(1-X)]}{1-q+\nu^2[X+q(1-X)]}.
    	\end{split}
    \end{equation*}
    This expression is formally identical to the composite-system stored energy formula in the bipartite Scenario 1. Its physical meaning is as follows: when the battery ground-state population is large, the external drive injects net energy into the composite system via the Unruh effect. When the excited-state population is large enough to make the numerator negative, de-excitation processes induced by the Unruh effect dominate, and the composite system instead loses energy.
    
    \subsection{Scenario 2}
    In this subsection, we discuss the case where the nearby charger $b$ is accelerated while the battery $a$ and the other charger $c$ remain static. Tracing over the composite-system density matrix $\varrho_{abc}^2$, the reduced density matrix of the battery subsystem is:
    \begin{small}
    	\begin{equation*}
    		\rho_{a}^2=\frac{1}{N_2}\left(
    		\begin{array}{cc}
    			\varrho_{11}+\varrho_{22}+\frac{\nu^2}{1-q}(\varrho_{11}+q\varrho_{22}) & 0\\
    			0 & (1+\frac{\nu^2}{1-q})\varrho_{44}\\
    		\end{array}
    		\right ).
    	\end{equation*}
    \end{small}
    Next, we compute the population difference between the excited and ground states of the battery subsystem:
    \begin{equation}
    	\begin{split}
    		P_e-P_g&=\frac{\varrho_{44}-\varrho_{11}-\varrho_{22}+\frac{\nu^2}{1-q}(\varrho_{44}-\varrho_{11}-q\varrho_{22})}{1+\frac{\nu^2}{1-q}(\varrho_{11}+\varrho_{44}+q\varrho_{22})}\\
    		&=\frac{2\varrho_{44}-1+\frac{\nu^2}{1-q}(\varrho_{44}-\varrho_{11}-q\varrho_{22})}{1+\frac{\nu^2}{1-q}(1-\varrho_{22}+q\varrho_{22})}
    	\end{split}
    \end{equation}
    This means that in three-body Scenario 2, the population difference $P_e- P_g$ depends on two parameters: the initial excitation probability of battery $a$, $\varrho_{44}$, and that of the accelerated charger $b$, $\varrho_{22}$.  
    
    Let $\eta=\frac{\nu^2}{1-q}$. According to Eq. (26), the numerator of the population difference of the battery subsystem is:
    \begin{equation}
    	F(q)=2\varrho_{44}-1+\eta(\varrho_{44}-\varrho_{11}-q\varrho_{22}).
    \end{equation} 
    The sudden emergence of ergotropy requires the following conditions:
    \begin{equation*}
    	\begin{split}
    		&F(0)<0,\,\text{i.e.}\,1-2\varrho_{44}>\nu^2(\varrho_{44}-\varrho_{11}),\\
    		&F(q_c)=0,\,0<q_c\leq0.4.
    	\end{split}
    \end{equation*}
    Now, by solving the equation $F(q_c)=0$ for $q_c$, we obtain:
    \begin{equation}
    	q_c=1-\frac{\nu^2(1-2\varrho_{44})}{\nu^2\varrho_{22}-(1-2\varrho_{44})}.
    \end{equation}
    
    The condition $F(0)<0$ can be further rewritten as 
    \begin{equation}
    \nu^2\varrho_{22}<(1-2\varrho_{44})(1+\nu^2).
    \end{equation}
    We now prove that $q_c>1$ under the condition of Eq. (29). If $\nu^2\varrho_{22}\leq(1-2\varrho_{44})$, then
    \begin{equation*}
    	q_c=1-\frac{\nu^2(1-2\varrho_{44})}{\nu^2\varrho_{22}-(1-2\varrho_{44})}\geq1.
    \end{equation*}
    If $\nu^2\varrho_{22}>(1-2\varrho_{44})$, then $q_c>0$ is equivalent to 
    	$\nu^2\varrho_{22}>(1-2\varrho_{44})(1+\nu^2)$, which contradicts Eq. (29).
    
   Over the entire parameter range satisfying the initial passive-state condition, the critical point $q_c$ is either negative (no physical critical point) or greater than 1 (outside the perturbative regime). Hence, within the strict perturbative regime $q\leq0.4$, there is no sudden emergence of ergotropy for the battery subsystem.
    \subsection{Scenario 3}
    In this subsection, we consider the scenario where both the battery and adjacent charger are acceleration simultaneously. Taking the partial trace over the final state, we obtain the ground-state population and the excited-state population of the battery subsystem as:
    \begin{equation*}
    	\begin{split}
    		&P_g=\frac{1}{N_3}(\frac{\nu^2q}{1-q}S+\varrho_{11}+\varrho_{22}+\frac{\nu^2}{1-q}\varrho_{11}),\\
    		&P_e=\frac{1}{N_3}[\varrho_{44}+\frac{\nu^2}{1-q}(1+\varrho_{24}+\varrho_{42})],
    	\end{split}
    \end{equation*}
    where $S=\varrho_{22}+\varrho_{44}+\varrho_{24}+\varrho_{42}$, and $N_3=1+\frac{\nu^2}{1-q}[2\varrho_{11}+(1+q)S]$. The population difference is:
    \begin{equation*}
    	\begin{split}
    		P_e-P_g&=\frac{\varrho_{44}-\varrho_{11}-\varrho_{22}+\nu^2S}{N_3}\\
    		&=\frac{2\varrho_{44}-1+\nu^2S}{1+\frac{\nu^2}{1-q}[2\varrho_{11}+(1+q)S]}.
    	\end{split}
    \end{equation*}
    Note that the numerator is entirely independent of the acceleration parameter $q$. In this case, $q$ merely adjusts the normalization factor, thereby affecting the ergotropy of the battery subsystem. Since $\frac{1-q}{1+q}$ is monotonically increasing on $q\in[0,0.4]$, the denominator increases monotonically with $q$. Based on this fact, we obtain:
    \\
    (i) If the numerator is nonpositive, then $P_e-P_g<0$ for all $q$, and the battery subsystem ergotropy exhibits no sudden emergence and remains identically zero.
    \\
    (ii) If the numerator is positive, then $P_e-P_g>0$ for all $q$, and the ergotropy of the battery subsystem decreases monotonically with $q$.
    
    This result is physically consistent with the conclusion in the bipartite co-acceleration case. The coherent superposition of multiple detectors coupled to the same quantum field cancels the $q$-dependence in the numerator, eliminating the pumping ability of the Unruh effect while retaining only the dilution effect. Hence, no sudden emergence of ergotropy occurs.
    
    For all three tripartite scenarios, we have verified that $\text{Tr}(\varrho_{abc}^{i=1,2,3})=1$ and $P_e+P_g=1$. The reduced battery populations are consistently obtained by grouping the basis states according to the battery being in $|g\rangle_a$ or $|e\rangle_a$.
     
   We briefly comment on the experimental relevance and limitations of our results. The Unruh effect typically requires extremely large accelerations to produce observable signatures in real detectors, which poses a practical challenge for direct experimental verification. However, the two-level Unruh-DeWitt detector employed here serves as a versatile theoretical tool that can be effectively simulated in analog platforms, such as trapped ions and superconducting circuits, where the effective acceleration is engineered through external drivings. Within this context, our predictions provide a qualitative guide for exploring how relativistic motion may influence quantum energy storage. We also emphasize that our analysis is restricted to the perturbative regime and does not cover the strong-coupling or ultra-relativistic limits. A full thermodynamic account including the work cost of maintaining acceleration, as well as the non-perturbative analysis of the $q\rightarrow1$ regime, remains open for future investigation.

	\section{IV. Conclusions and discussions}
  
  We have systematically investigated the effects of uniform acceleration on the ergotropy of quantum batteries in both bipartite and tripartite setups, with the accelerating party modeled as a point-like Unruh-DeWitt detector. In the bipartite system, three scenarios were analyzed. Accelerating the battery induces a sudden emergence of ergotropy at a critical threshold, accompanied by net energy injection into the composite system. Accelerating the charger leaves the battery ergotropy constant within the perturbative regime, zero if initially passive, constant if initially active, as the $q$-dependence is suppressed to $O(\nu^4)$. Simultaneous acceleration of both subsystems leads to monotonic decay or identically zero ergotropy, due to coherent cancellation of the $q$-dependence in the numerator.
  
  Extending to a tripartite system with one battery and two chargers, we found that battery acceleration reproduces the bipartite emergence behavior, while accelerating either charger does not induce ergotropy emergence—consistent with the bipartite charger-acceleration case. Simultaneous acceleration in the tripartite setting again eliminates sudden emergence, consistent with the bipartite counterpart.

  Beyond the present analysis, our results suggest that Unruh-induced population inversion could serve as a viable mechanism for charging quantum batteries in relativistic settings. While the energy pumped into the battery ultimately originates from the external agent maintaining the acceleration, the Unruh effect acts as a catalyst that redistributes this energy into the extractable form. This opens up the possibility of controlling battery performance through motional states, analogous to how external fields are used in conventional charging protocols.

	\bigskip
	{\bf Acknowledgments:} ~This work is supported by the National Natural Science Foundation of China (NSFC) under Grant No.12564048; the Natural Science Foundation of Hainan Province under Grant No. 125RC744 and the specific research fund of the Innovation Platform for Academicians of Hainan Province.

	\section{Appendix A: Unruh Effect on Tripartite Quantum Batteries}
	\setcounter{equation}{0}
	\renewcommand{\theequation}{A\arabic{equation}}
	Since the interaction term given by system Hamiltonian ensures the conservation of the total system excitations, the evolution of the system involves only the three states $|gge\rangle, |geg\rangle$, and $|egg\rangle$. Thus, we can restrict the dynamical evolution to a three-dimensional subspace, and the total system Hamiltonian $H$ is simplified accordingly. Specifically, 
   \begin{equation*}
   	\begin{split}
   		\langle gge|H|gge\rangle&=\langle gge|(\kappa_ac_a^\dagger c_a+\kappa_bc_b^\dagger c_b+\kappa_cc_c^\dagger c_c)|gge\rangle\\
   		&+J_1\langle gge|(c_a^\dagger c_b+c_ac_b^\dagger)|gge\rangle\\
   		&+J_2\langle gge|(c_b^\dagger c_c+c_bc_c^\dagger)|gge\rangle\\
   		&=(0+0+\kappa)+0+0\\
   		&=\kappa.
   	\end{split}
   \end{equation*}
   In the equation above, we have used the properties of the creation and annihilation operators: $c_{a(bc)}|g\rangle_{a(bc)}=c_{a(bc)}^\dagger|e\rangle_{a(bc)}=0$, $c_{a(bc)}|e\rangle_{a(bc)}=|g\rangle_{a(bc)}$, and $c_{a(bc)}^\dagger|g\rangle_{a(bc)}=|e\rangle_{a(bc)}$.
   \begin{equation*}
   	\begin{split}
   		\langle gge|H|geg\rangle&=\langle gge|(\kappa_ac_a^\dagger c_a+\kappa_bc_b^\dagger c_b+\kappa_cc_c^\dagger c_c)|geg\rangle\\
   		&+J_1\langle gge|(c_a^\dagger c_b+c_ac_b^\dagger)|geg\rangle\\
   		&+J_2\langle gge|(c_b^\dagger c_c+c_bc_c^\dagger)|geg\rangle\\
   		&=(0+0+0)+0+J_2\\
   		&=J_2.
   	\end{split}
   \end{equation*}
   \begin{equation*}
   	\begin{split}
   		\langle gge|H|egg\rangle&=\langle gge|(\kappa_ac_a^\dagger c_a+\kappa_bc_b^\dagger c_b+\kappa_cc_c^\dagger c_c)|egg\rangle\\
   		&+J_1\langle gge|(c_a^\dagger c_b+c_ac_b^\dagger)|egg\rangle\\
   		&+J_2\langle gge|(c_b^\dagger c_c+c_bc_c^\dagger)|egg\rangle\\
   		&=(0+0+0)+0+0\\
   		&=0.
   	\end{split}
   \end{equation*}

   \begin{equation*}
   	\begin{split}
   		\langle geg|H|egg\rangle&=\langle geg|(\kappa_ac_a^\dagger c_a+\kappa_bc_b^\dagger c_b+\kappa_cc_c^\dagger c_c)|egg\rangle\\
   		&+J_1\langle geg|(c_a^\dagger c_b+c_ac_b^\dagger)|egg\rangle\\
   		&+J_2\langle geg|(c_b^\dagger c_c+c_bc_c^\dagger)|egg\rangle\\
   		&=(0+0+0)+J_1+0\\
   		&=J_1.
   	\end{split}
   \end{equation*}
   Repeating this process, the system Hamiltonian can be rewritten as
   \begin{equation*}
   	H=\left(
   	\begin{array}{ccc}
   		\kappa & J_2 & 0\\
   		J_2 & \kappa & J_1\\
   		0 & J_1 & \kappa\\
   	\end{array}
   	\right ).
   \end{equation*}
   The eigenvalues of the Hamiltonian are $e_1=\kappa$, $e_2=\kappa-\sqrt{J_1^2+J_2^2}$, $e_3=\kappa+\sqrt{J_1^2+J_2^2}$, with the corresponding eigenvectors are
   \begin{small}
   	\begin{equation*}
   		\begin{split}
   			|e_1\rangle&=-\frac{J_1}{\sqrt{J_1^2+J_2^2}}|gge\rangle+\frac{J_2}{\sqrt{J_1^2+J_2^2}}|egg\rangle,\\
   			|e_2\rangle&=\frac{J_2}{\sqrt{2(J_1^2+J_2^2)}}|gge\rangle-\frac{1}{\sqrt{2}}|geg\rangle+\frac{J_1}{\sqrt{2(J_1^2+J_2^2)}}|egg\rangle,\\
   			|e_3\rangle&=\frac{J_2}{\sqrt{2(J_1^2+J_2^2)}}|gge\rangle+\frac{1}{\sqrt{2}}|geg\rangle+\frac{J_1}{\sqrt{2(J_1^2+J_2^2)}}|egg\rangle.
   		\end{split}
   	\end{equation*}
   \end{small}
   We express the initial state $|gge\rangle$ in terms of the eigenvectors $|e_1\rangle$, $|e_2\rangle$, and $|e_3\rangle$ as:
   \begin{small}
   	\begin{equation*}
   		|gge\rangle=\frac{-J_1}{\sqrt{J_1^2+J_2^2}}|e_1\rangle+\frac{J_2}{\sqrt{2(J_1^2+J_2^2)}}|e_2\rangle+\frac{J_2}{\sqrt{2(J_1^2+J_2^2)}}|e_3\rangle.
   	\end{equation*}
   \end{small}
   Thus, the state of the total system at time $t$ can be written as:
   \begin{small}
   	\begin{equation*}
   		\begin{split}
   			|\psi(t)\rangle_{abc}&=\exp(-iHt)|gge\rangle\\
   			&=\frac{-J_1e^{-ie_1t}}{\sqrt{J_1^2+J_2^2}}|e_1\rangle+\frac{J_2e^{-ie_2t}}{\sqrt{2(J_1^2+J_2^2)}}|e_2\rangle+\frac{J_2e^{-ie_3t}}{\sqrt{2(J_1^2+J_2^2)}}|e_3\rangle\\
   			&=\alpha_1(t)|gge\rangle+\alpha_2(t)|geg\rangle+\alpha_3(t)|egg\rangle.
   		\end{split}
   	\end{equation*}
   \end{small}
   
\end{document}